\documentclass[reprint,a4paper,twocolumn,aps,pra,amsmath,amssymb,floatfix,showkeys,longbibliography
]{revtex4-2}

\usepackage{color}
\usepackage{dcolumn}
\usepackage{graphicx}

\usepackage{amsmath,amsfonts,amssymb,amsthm}
\usepackage{mathtools}

\usepackage{bbold}
\usepackage{dsfont}

\usepackage{soul}

\usepackage{fancyhdr}
\usepackage{multirow}
\usepackage{bm}
\usepackage{dsfont}
\usepackage{cancel}

\usepackage{xifthen}

\usepackage{float}
\graphicspath{{"./graphs/"}{"./scheme/"}}

\usepackage{graphicx}
\usepackage[usenames,dvipsnames,dvipsnames,svgnames,table]{xcolor}
\usepackage[colorlinks,bookmarks=false,
linkcolor=blue,
urlcolor=blue,
citecolor=blue
]{hyperref}

\usepackage{xspace}

\renewcommand{\L}{\mathcal{L}}

\usepackage{physics}

\usepackage{xifthen}

\def\op#1{\hat{#1}}
\renewcommand{\ao}[1][]{%
	\ifthenelse{\equal{#1}{}}{\ensuremath{\op{a}}}{\ensuremath{\op{#1}}}%
}

\newcommand{\co}[1][]{%
	\ifthenelse{\equal{#1}{}}{\ensuremath{{{}\op{a}^{\dagger}}}}{\ensuremath{{{}\op{#1}^{\dagger}}}}%
}

\makeatletter
\newcommand*{\transpose}{\bgroup\@transpose}
\newcommand*{\@transpose}[1][0]{\mathpalette\@@transpose{#1}\egroup}
\newcommand*{\@@transpose}[2]{\setbox0=\hbox{\m@th$#1\mkern-#2mu\intercal$}\raise\dp0\box0}
\makeatother

\usepackage{xcolor,colortbl,mathdots}

\usepackage{makecell}

\usepackage{enumitem} 
\usepackage{array,calc}

\usepackage[english]{babel}
\usepackage{pgfplots} 
\pgfplotsset{compat=1.18}
\usetikzlibrary{plotmarks}

\usepackage{scalerel}

\usepackage[normalem]{ulem}

\def\ext{0}
\usepackage{ifthen}
\usepackage{xifthen}
\newcommand{\comments}[1]{
	\ifthenelse{\equal{\ext}{1}}{\textcolor{blue}{#1}}{}
}
\def\extFuture{1}
\definecolor{darkgreen}{RGB}{0,140,0}

\definecolor{rowgray}{gray}{0.93}

\graphicspath{{"../graphs/"}{"../images/"}{"../scheme/"}{"figures/"}
}

\usepackage{placeins}
\begin{document}

\title{Generalized master equation for driven quantum oscillators:\\
microscopic origin of nonlinear dissipation and asymmetric resonances}
\author{Jakob Wagner}
\affiliation{Department of Physics, University of Konstanz, 78464 Konstanz, Germany}
\author{Jeff Maki}
\affiliation{Department of Physics, University of Konstanz, 78464 Konstanz, Germany}
\author{Oded Zilberberg}
\affiliation{Department of Physics, University of Konstanz, 78464 Konstanz, Germany}
\author{Kilian Seibold}
\affiliation{Department of Physics, University of Konstanz, 78464 Konstanz, Germany}

\date{\today}

\begin{abstract}
Driven nonlinear quantum oscillators are a central platform for quantum technologies, yet their dissipative dynamics are typically described using Lindblad or Caldeira-Leggett master equations derived under assumptions that exclude nonlinearities and driving.
Here, we derive a generalized Caldeira-Leggett master equation for driven nonlinear oscillators by retaining the full nonlinear and time-dependent system dynamics in the construction of the dissipator.
For position- and momentum-dependent system-bath coupling, the dissipator itself becomes dynamically dressed, generating nonlinear and drive-dependent dissipative channels beyond conventional fixed-dissipator approaches.
This produces nonlinear damping together with dissipation-induced corrections to the effective drive.
The resulting dissipative dynamics suppress large-amplitude excitations and reduce phase-space fluctuations.
For a driven Kerr oscillator, this leads to the suppression of bistability, asymmetric resonance responses, and strongly modified fluctuation distributions.
More broadly, our results establish a microscopic framework in which nonlinear dynamics and driving directly reshape the dissipative sector of driven open quantum systems.
\end{abstract}

\maketitle


\section{Introduction}
Driven open quantum systems play a central role in modern quantum technologies~\cite{BreuerPetruccione2002, Nielsen2010, Wiseman2010, Gardiner2004}. 
Their rich behavior arises from the interplay of coherent nonlinear dynamics~\cite{T_rschmann_2019}, external periodic driving~\cite{Herr2012, Goldman2014, Oka2019, Rudner2020}, and engineered dissipation~\cite{Poyatos1996}, enabling applications such as quantum state preparation~\cite{Poyatos1996, Verstraete2009, Shankar2013, Leghtas2015}, sensing~\cite{DiCandia2023, Heugel2019}, precision metrology~\cite{Ilias2022}, quantum simulation~\cite{Gelin2021}, and nonequilibrium phases of matter~\cite{Fitzpatrick2017}.
A standard theoretical framework is the Lindblad-Gorini-Kossakowski-Sudarshan (LGKS) master equation~\cite{Lindblad1976, Chruscinski2017}, which provides a time-local, trace-preserving, and completely positive description of the reduced dynamics.
Microscopically, LGKS-type equations are typically derived from system-bath models such as the Caldeira-Leggett framework, where the system is linearly coupled to a bath of harmonic oscillators via its canonical position quadrature~\cite{Caldeira1983, Caldeira1983a, BreuerPetruccione2002, Schlosshauer2008}. 
These derivations rely on restrictive assumptions including weak system-bath coupling, short bath memory, harmonic or linearized dynamics, and weak or slowly varying driving.
As a result, the dissipator is constructed using harmonic dynamics, such that nonlinearities and explicit time dependence enter only the unitary evolution while the dissipative sector remains effectively linear and time independent.

These assumptions are routinely violated in contemporary experiments.
Platforms such as superconducting SQUIDs~\cite{Venkatraman2023, Venkatraman2024, Duffus2016, Duffus2017, Duffus2018}, nanomechanical devices~\cite{Ochs_2022, Eichler2011}, and cavity-QED systems~\cite{Walls2008} operate in regimes with strong nonlinearities and large-amplitude driving.
In many of these systems, dissipation can involve both position- and momentum-type coupling channels, corresponding in superconducting circuits to interactions mediated by flux and charge degrees of freedom~\cite{Krantz2019, Blais2021}.
Momentum-coupled open-system dynamics and generalized dissipation models have been investigated in a variety of contexts~\cite{Ullersma1966, Walls1970, Huang2022, Venkatraman2024, bhattacharjee2024, Bai2005, Duffus2016, Duffus2017, Duffus2018, Gelin2021}.
Standard LGKS descriptions often fail in these regimes~\cite{Eichler2011, Venkatraman2023, Ochs_2022, Duffus2016, Duffus2017, Duffus2018}, while previous approaches typically treated momentum coupling, nonlinearities, and driving separately~\cite{Ullersma1966, Caldeira1983, Caldeira1983a, Grabert1988, Hu1992, Isar_1994, Zerbe1995, Peano2004, Grabert2018, Dann2018, Lampo2016, Vacchini2002, Banerjee2003, Gao1997, Duffus2016, Duffus2017, Duffus2018, Huang2022, bhattacharjee2024}.
This calls for a unified framework incorporating general system-bath coupling together with nonlinear dynamics and time-dependent driving directly in the dissipative evolution.

In this work, we derive a generalized Caldeira-Leggett master equation for a nonlinear periodically driven oscillator by retaining the full nonlinear and time-dependent system dynamics throughout the dissipative evolution.
As a result, the dissipator is no longer fixed solely by the bare canonical variables, but dynamically inherits the nonlinear and driven operator structure of the system Hamiltonian.
As a consequence, even bilinear system-bath coupling generates nonlinear and drive-dependent dissipative processes beyond standard Lindblad and Caldeira-Leggett descriptions.
This naturally gives rise to effects such as dissipation-induced squeezing~\cite{Duffus2017, Rastelli2016}, nonlinear damping~\cite{Eichler2011}, and dissipation-induced corrections to the effective drive~\cite{Peano2004, Grabert1988, Grabert2018, Zerbe1995, Dann2018}, as well as modified fluctuation relations and decay rates~\cite{Venkatraman2023, Caldeira1983, Caldeira1983a, Huang2022, Hu1993, Diosi1993, Supplemental}.
Previous approaches typically introduced nonlinear damping phenomenologically, while drive-induced dissipative effects were treated numerically.
We apply the framework to a driven Kerr oscillator.
Our approach predicts amplitude-dependent damping, dissipation-induced driving, suppression of bistability, and modified fluctuation distributions, providing clear experimental signatures of momentum-mediated dissipation.
Overall, our results establish a microscopically grounded framework for driven nonlinear open quantum systems beyond linear-response and static-dissipation regimes.

The paper is organized as follows.
In Sec.~\ref{sec: Theoretical framework}, we introduce the driven nonlinear oscillator model, derive the generalized Caldeira-Leggett master equation, and discuss its key properties in Sec.~\ref{subsec: discussion}.
In Sec.~\ref{subsec: nonlin_damping_cl} and~\ref{subsec: nonlin_damping_qm}, we apply the formalism to a Kerr nonlinear oscillator and analyze classical dynamics, quantum steady-states, occupation probabilities, and nonlinear damping.
Sections~\ref{subsec: driven_system} and~\ref{subsec: MPR} address the driven regime, including dissipation-induced driving, multiphoton-resonance phenomena, and fluctuation reshaping.
Finally, we conclude with an outlook and potential applications in Sec.~\ref{sec: conclusion}.



\section{Model and Derivation}
\label{sec: Theoretical framework}

\subsection{Overview}

In the following, we derive the generalized master equation in a time-local Born-Markov treatment without resorting to linear system dynamics.
Our starting point is a driven nonlinear quantum oscillator with a microscopic system-bath Hamiltonian with both position-position and momentum-momentum coupling, see Fig.~\ref{fig: system}.
The derivation differs from standard treatments in one key aspect: we retain the full nonlinear and explicitly time-dependent system Hamiltonian throughout the derivation.
As a consequence, the dissipative channels describing the system-bath interaction inherit the nonlinear and driven dynamics of the isolated system.
This leads to a dynamical dressing of the system-bath coupling, through which intrinsic nonlinearities generate amplitude-dependent dissipation, while external driving produces dissipation-induced corrections to the effective drive amplitude and phase.

The resulting generalized Caldeira-Leggett master equation is trace preserving and Hermiticity preserving, but generally not completely positive outside specific regimes.
It should therefore be viewed as an effective description valid for weak system-bath coupling, short bath correlation times, and sufficiently high temperatures.
Within this regime, it captures nonlinear and drive-dependent modifications of the dissipative dynamics that are absent in fully secularized Lindblad approaches.

\subsection{General Hamiltonian}

\begin{figure}[t]
\centering
\includegraphics[width=0.45\textwidth]{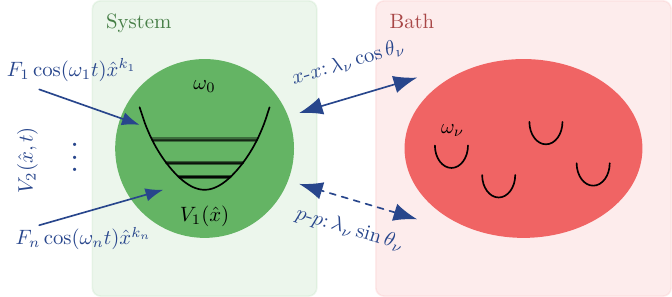}
\caption{
\textit{Driven nonlinear oscillator coupled to a bosonic environment.} A quantum oscillator of frequency $\omega_0$ with intrinsic nonlinearity $V_1(\op{x})$ is subject to time-dependent driving $V_2(\op{x},t)$ [cf.~Eq.~\eqref{eq:H_syst}]. 
The oscillator is coupled to a bath of bosonic resonators, $\nu$, with frequencies $\omega_\nu$. 
The system-bath interaction [cf.~Eq.~\eqref{eq: H_Int}] interpolates continuously between two limits: purely position-based coupling ($\op{x}\op{x}_{\nu}$) at $\theta=0$, and purely momentum-based coupling ($\op{p}\op{p}_{\nu}$) at $\theta=\pi/2$.
}
\label{fig: system}
\end{figure}

We consider a driven nonlinear quantum oscillator coupled to a bosonic bath (Fig.~\ref{fig: system}).
The total Hamiltonian reads
\begin{equation}
    \op{H}=\op{H}_S(t)+\op{H}_B+\op{H}_{SB}\;,
    \label{eq: Hamiltonian system + bath + interaction}
\end{equation}
where $\op{H}_S(t)$ is the Hamiltonian for the system, $\op{H}_B$ the bath, and $\op{H}_{SB}$ the system-bath interaction.

The system Hamiltonian is given by
\begin{equation}
\op{H}_{S}(t) = \frac{\op{p}^2}{2m} + \frac{1}{2}m\omega_0^2\op{x}^2 + V(\op{x},t)\;,
\label{eq:H_syst}
\end{equation}
with canonical position and momentum operators $\op{x}$ and $\op{p}$, mass $m$, and bare frequency $\omega_0$.
The potential is decomposed as
$V(\op{x},t)=V_1(\op{x})+V_2(\op{x},t)$,
where $V_1$ describes intrinsic nonlinearities and $V_2$ external driving.
We take $V_1(\op{x})=\sum_n \tfrac{g_n}{n}\op{x}^n$ with coefficients $g_n\in \mathbb{R}$ for static anharmonic corrections to the potential, and
$V_2(\op{x},t)=\sum_n F_n\cos(\omega_n t)\op{x}^{k_n}$, for periodic driving terms with amplitude $F_n\in \mathbb{R}$, frequency $\omega_n$, and polynomial order $k_n\in\mathbb{N}$.
This form treats anharmonicity and driving on equal footing and allows us to track their impact on the dissipative dynamics.

We model the bath as a collection of independent harmonic oscillators labeled by $\nu$, with canonical position and momentum operators $\op{x}_{\nu}$ and $\op{p}_{\nu}$, masses $m_{\nu}$, and frequencies $\omega_{\nu}$,
\begin{equation}
    \op{H}_B = \sum_\nu \left[ \frac{\op{p}_{\nu}^2}{2m_\nu} + \frac{1}{2}m_\nu \omega_{\nu}^2 \op{x}_{\nu}^2 \right]\;,
\end{equation}
which provides a standard microscopic description of the environment~\cite{Caldeira1983, Caldeira1983a, BreuerPetruccione2002}.
The system-bath interaction is taken bilinear in the system and bath position and momentum operators~\footnote{%
The frequency factors ensure consistent dimensions of the position and momentum coupling terms.},
\begin{equation}
    \op{H}_{SB} = \sum_\nu\lambda_\nu\left[
    \cos(\theta_\nu)\op{x}\op{x}_\nu
    + \frac{\sin(\theta_\nu)}{m m_\nu \omega_0 \omega_\nu}\,\op{p}\op{p}_\nu
    \right]\;,
    \label{eq: H_Int}
\end{equation}
with $\lambda_{\nu}$ the coupling strength to the bath mode $\nu$.
This form generalizes the standard Caldeira-Leggett model by incorporating momentum-momentum coupling in addition to position-position coupling, and is essential for capturing the amplitude-dependent damping and dissipative drive corrections discussed below.
The mixing angle $\theta_\nu \in[0,\pi/2]$ controls the relative weight between the two coupling channels.

The limits $\theta_\nu=0$ and $\theta_\nu=\pi/2$ correspond to purely position-position ($x$-$x$)~\cite{Caldeira1983, Caldeira1983a, BreuerPetruccione2002, Schlosshauer2008} and momentum-momentum ($p$-$p$) coupling~\cite{Gelin2021, Venkatraman2024, Cuccoli2001}, respectively, while $\theta_\nu=\pi/4$ yields equal contributions.
For simplicity, we assume a mode-independent mixing angle $\theta$.
At $\theta=\pi/4$, the dissipative sector reduces to the structure of a thermal Lindblad master equation, up to a trivial prefactor~\cite{Duffus2016, Duffus2017, Bai2005, bhattacharjee2024, Huang2022, Duffus2018, Walls1970}.
Away from this Lindblad point, the dynamics is of Redfield type and not guaranteed to be completely positive.
This interpolation provides a controlled connection between Caldeira-Leggett and Lindblad limits while retaining additional dynamical corrections arising from momentum-mediated dissipation.


\subsection{Generalized master equation}

We derive a time-local master equation for the reduced dynamics that incorporates intrinsic nonlinearities, time-dependent driving, and mixed position-position and momentum-momentum system-bath coupling on equal footing within the dissipative dynamics.

We start from the von Neumann equation for the total density matrix of the system and bath, $\op{\rho}_{\text{tot}}$,
\begin{equation}
    \partial_t \op{\rho}_{\text{tot}}(t) = \frac{1}{\mathrm{i}\hbar}\left[\op{H},\op{\rho}_{\text{tot}}(t)\right]\;,
\end{equation}
with $\op{H}$ defined in Eq.~\eqref{eq: Hamiltonian system + bath + interaction}. 
We treat the system-bath coupling by moving to the interaction picture with respect to the uncoupled system and bath dynamics.
The interaction-picture representation of an operator $\op{O}(t)$ reads \begin{equation}
    \op{O}_{I}(t,t_0)=\op{\mathcal{U}}^\dagger(t,t_0)\op{O}(t)\op{\mathcal{U}}(t,t_0),
\end{equation}
with evolution operator
\begin{equation}
    \op{\mathcal{U}}(t,t_0)=\mathcal{T}\exp\!\left[\frac{1}{\mathrm{i}\hbar}\int_{t_0}^t dt\,(\op{H}_S(t)+\op{H}_B)\right]\;,
\end{equation}
where $\mathcal{T}$ denotes time ordering. The density matrix then evolves as
\begin{equation}
    \partial_t\op{\rho}_{\text{tot,}I}(t,t_0)=\frac{1}{\mathrm{i}\hbar}[\op{H}_{SB,I}(t,t_0), \op{\rho}_{\text{tot,}I}(t,t_0)]\;.
    \label{eq:vN_int_pic}
\end{equation}
Crucially, we retain the full nonlinear and time-dependent structure of $\op{H}_S(t)$ when constructing the interaction-picture operators~\cite{Peano2004, Duffus2016, Duffus2017, Duffus2018}.
This contrasts with standard treatments~\cite{BreuerPetruccione2002, Schlosshauer2008}, where the system dynamics is typically linearized or treated perturbatively prior to evaluating the dissipative kernel.

We evaluate the system-bath coupling within the Born-Markov approximation~\cite{BreuerPetruccione2002, Schlosshauer2008}.
The Born approximation assumes weak system-bath coupling, such that the total density matrix allows $\op{\rho}_{\text{tot}}(t)\approx\op{\rho}(t)\otimes\op{\rho}_B$, where $\op{\rho}_B$ is the stationary equilibrium state of the bath. 
Meanwhile the Markov approximation neglects bath memory effects and leads to time-local dynamics, allowing us to insert the formal solution of Eq.~\eqref{eq:vN_int_pic} in itself and evaluate the density matrix at times $(t, t_0)$, while also extending the integration limits:
\begin{equation}
\begin{aligned}
    &\partial_t\op\rho_{\text{tot,}I}(t, t_0)
    \\
    =& \frac{1}{\mathrm{i}\hbar}[\op H_{SB,I}(t, t_0), \op\rho_{\text{tot,}I}(t_0, t_0)]-\frac{1}{\hbar^2}\int_{0}^\infty d\tau\,
    \\
    &\times\!\left(
    [\op{H}_{SB, I}(t, t_0),
    [\op{H}_{SB, I}(t-\tau, t_0),
    \op{\rho}_{I}(t, t_0)\otimes\op{\rho}_{B}]]
    \right)
    \\
    &+\mathcal{O}(\hbar^{-3})\;.
\end{aligned}
\label{eq:vN_int_pic_exp}
\end{equation}
Tracing over the bath degrees of freedom in Eq.~\eqref{eq:vN_int_pic_exp}, while assuming $\mathrm{Tr}_B\left([\op H_{SB,I}(t, t_0), \op\rho_{\text{tot,}I}(t_0, t_0)]\right)=0$ 
and transforming back to the Schrödinger picture yields a time-local Redfield-type master equation,
\begin{equation}
\begin{aligned}
\partial_t\op{\rho}(t)
=& \frac{1}{\mathrm{i}\hbar}
 [\op{H}_S(t),\op{\rho}(t)]
 -\frac{1}{\hbar^2}
 \int_0^\infty \! d\tau\;
\\
&\times\,
 \mathrm{Tr}_B\!\Big(
   [\op{H}_{SB},
     [\op{H}_{SB,I}(t-\tau,t),
      \op{\rho}(t)\!\otimes\!\op{\rho}_B]]
 \Big)
\\
&+\mathcal{O}(\hbar^{-3}) \; .
\end{aligned}
\label{eq:first_order_dm}
\end{equation}
The $\mathcal{O}(\hbar^{-1})$ contribution is simply the unitary time evolution of the system's density matrix due to the system Hamiltonian.
The $\mathcal{O}(\hbar^{-2})$ term, on the other hand, describes the dissipative backaction of the bath. 

This term in the master equation is determined by two ingredients:
(i) the time evolution of system operators in the interaction picture, and
(ii) the bath correlation functions.
While in previous works, the nonlinearity of the system and the drive is only kept in the $\mathcal{O}(\hbar^{-1})$ term of the master equation, i.e.\ the unitary dynamics generated by the system Hamiltonian, here we keep such terms in the interaction picture of the system-bath coupling  $H_{SB,I}(t,t_0)$.
Consequently, the system-bath coupling becomes dynamically dressed, leading to amplitude-dependent dissipative corrections from intrinsic nonlinearities and dissipation-induced corrections to the effective drive amplitude and phase from external driving.
In the generalized framework, the dissipator is therefore no longer fixed solely by the bare canonical variables, but dynamically inherits the nonlinear and driven operator structure of the system Hamiltonian. As a result, nonlinearities and external driving modify not only the unitary evolution, but also the dissipative channels themselves.

%
We evaluate the bath correlators for an Ohmic bath with a Lorentz-Drude spectral cutoff frequency $\Omega$~\cite{Caldeira1983, Caldeira1983a}. 
We assume $\hbar\Omega$ to be the largest energy scale, such that $\hbar\omega_0$, $g_n(\hbar/m\omega_0)^{n/2}$, and $F_n(\hbar/m\omega_0)^{k_n/2}$ are all $\ll\hbar\Omega$.
Under these conditions, the bath correlation function  in Eq.~\eqref{eq:first_order_dm} is sharply peaked at short times, with a memory time $\sim \hbar/\Omega$.
This justifies a short-time expansion of the interaction-picture system operators, which we truncate at first order in time~\cite{Duffus2016, Duffus2017, Duffus2018}. 
Physically, this implies that the bath responds effectively instantaneously to the nonlinear and driven system dynamics, allowing these dynamics to modify both the effective damping and the effective drive.

In order to explicitly evaluate Eq.~\eqref{eq:first_order_dm}, we assume that the system-bath spectral density is strongly peaked near the dominant transition frequency $\omega_0$ of the driven nonlinear oscillator. As shown in the Supplemental Material~\cite{Supplemental}, this procedure yields a generalized time-local master equation 
\begin{equation}
    \partial_t \op{\rho}(t)=\mathcal{L}(t)\op{\rho}(t)\;,
\end{equation}
with the time-dependent Liouvillian  $\mathcal{L}(t)$ explicitly reflecting the nonlinear and driven system dynamics
\begin{widetext}
    \begin{subequations}
    \label{eq: improved_Liouvillian}
    \begin{align}
    \L(t)\op{\rho} =&\frac{1}{i\hbar}\left[\op{H}_S(t),\op{\rho}(t)\right]\label{eq:gCL_a}
    \\
        -&\left(1+\cos(2\theta)+\sin(2\theta)\right)\frac{\gamma m\omega_0}{4\hbar}c(T)[\op{x},[\op{x},\op{\rho}(t)]]-\left(1-\cos(2\theta)+\sin(2\theta)\right)\frac{\gamma}{4m\hbar\omega_0}c(T)[\op{p},[\op{p},\op{\rho}(t)]]\label{eq:gCL_b}
        \\
        -&\mathrm{i}\left(1+\cos(2\theta)+\sin(2\theta)\right)\frac{\gamma}{4\hbar}[\op{x},\{\op{p},\op{\rho}(t)\}]+\mathrm{i}\left(1-\cos(2\theta)+\sin(2\theta)\right)\frac{\gamma}{4\hbar}[\op{p}, \{\op{x}, \op{\rho}(t)\}]\label{eq:gCL_c}
        \\
        +&\frac{\gamma}{4\hbar^2}\left(\left(1-\cos(2\theta)\right)\frac{1}{m\omega_0^2}[\op{p}, \{[V_1(\op{x}),\op{p}],\op{\rho}(t)\}]+\mathrm{i}\sin(2\theta)\frac{1}{\omega_0}c(T)[\op{x}, [[V_1(\op{x}),\op{p}],\op{\rho}(t)]]\right)\label{eq:gCL_d}
        \\
        +&\sum_n\frac{\mathrm{i}\gamma k_nF_n\cos(\omega_n t)}{4\hbar}\left(\left(1-\cos(2\theta)\right)\frac{1}{m\omega_0^2}[\op{p}, \{\op{x}^{k_n-1},\op{\rho}(t)\}]+\mathrm{i}\sin(2\theta)\frac{1}{\omega_0}c(T)[\op{x}, [\op{x}^{k_n-1},\op{\rho}(t)]]\right)\;,\label{eq:gCL_e}
\end{align}
\end{subequations}
\end{widetext}
where $\gamma\sim\lambda_\nu^2/m_\nu\omega_\nu$ denotes the dissipation rate, and $c(T) = \coth\big(\hbar\omega_0/(2k_B T)\big)=2n_{\mathrm{th}}+1$ encodes the thermal occupation $n_{\mathrm{th}}$ of the bath modes at temperature $T$. 
Different conventions for evaluating bath correlation functions can lead to discrepancies by factors of two~\cite{BreuerPetruccione2002, Schlosshauer2008, Gao1997, Duffus2018, Dodonov2007, Diósi2012, Isar_1994}.
Here, we adopt a convention that reproduces the Bose-Einstein distribution and yields the correct classical limit for $\theta=0$.
Further details of the derivation and underlying assumptions are provided in the Supplemental Material~\cite{Supplemental}.


\subsection{Discussion}
\label{subsec: discussion}

Equation~\eqref{eq: improved_Liouvillian} defines the generalized Caldeira-Leggett master equation, gCL($\theta$), which constitutes the central result of this work.
Eq.~\eqref{eq: improved_Liouvillian} contains several terms which describe previous treatments of the Caldeira-Leggett model, as well as additional terms. First consider, Eq.~\eqref{eq:gCL_a}. This term describes the unitary von Neumann evolution generated by the system Hamiltonian defined in Eq.~\eqref{eq:H_syst}. The remaining terms are commonly denoted as the dissipator.
Bath-induced renormalizations of the oscillator mass and frequency (Lamb shifts)~\cite{Grabert2018, Kohler_1997, Karrlein1997, Zerbe1995, Mulder2021, Tanimura1991, Peano2004} are neglected in this work.

Equations~\eqref{eq:gCL_b} and~\eqref{eq:gCL_c} describe the decoherence and dissipation induced by the system-bath coupling.
In the limit $\theta=0$, the master equation reduces to the standard Caldeira-Leggett form associated with purely $x$-$x$ coupling~\cite{Caldeira1983, Caldeira1983a, BreuerPetruccione2002}.
This case, corresponding to mechanical damping, yields the correct classical limit~\cite{Dekker1977, Isar_1994}. To see this, we note that Eq.~\eqref{eq:gCL_c} can be decomposed into two distinct parts~\cite{Supplemental}: a purely dissipative contribution and an effective Hamiltonian term of the form
\begin{equation}
    \Lambda(\theta)=\frac{\gamma\cos(2\theta)}{4}\{\op{x},\op{p}\}\;.
\end{equation}
This term corresponds to a squeezing-like contribution ~\cite{Duffus2016, Duffus2017, Duffus2018, Dodonov2007} which is crucial for obtaining the correct equations of motion by breaking the position-momentum symmetry~\cite{Supplemental}.
In this $x$-$x$ coupling case, the position diffusion term $\propto [\op p, [\op p, \op\rho]]$ is typically introduced phenomenologically to ensure physical consistency~\cite{BreuerPetruccione2002, Christie_2024}.
Within the present framework, position diffusion and related dissipative contributions arise directly from the $p$-$p$ coupling.
Related diffusion terms have been identified in previous studies incorporating $p$-$p$ coupling~\cite{Duffus2016, Duffus2018, Gao1997, Kohler_1997, Duffus2017, Dekker1977, Diosi1993, Sandulescu1987, Isar_1994}.

Eqs.~\eqref{eq:gCL_a}-\eqref{eq:gCL_c}, interpolate between the standard Caldeira-Leggett equation and a Lindblad-type master equation. We denote the Caldeira-Leggett master equation formed from these three terms as CL($\theta$).
At $\theta=\pi/4$, the CL($\theta$) model has equal contributions from the position and momentum system-bath coupling, and the dissipator acquires a structure that can be mapped directly onto a Lindblad form (up to an overall prefactor) with both position and momentum loss channels~\cite{BreuerPetruccione2002, Schlosshauer2008, Dodonov2007, Isar_1994,Ferialdi2017}.
The corresponding Lindblad master equation reads
\begin{align}
\label{eq: standard_lindblad_ME}
    \partial_t\op{\rho}(t)=&\frac{1}{\mathrm{i}\hbar}[\op{H}_S(t),\op{\rho}(t)]
    \\
    &-\frac{\gamma m\omega_0}{4\hbar}c(T)[\op{x},[\op{x},\op{\rho}(t)]]
     -\frac{\gamma}{4m\hbar\omega_0}c(T)[\op{p},[\op{p},\op{\rho}(t)]]\nonumber
    \\
    &-\frac{\mathrm{i}\gamma}{4\hbar}[\op{x},\{\op{p},\op{\rho}(t)\}]
     +\frac{\mathrm{i}\gamma}{4\hbar}[\op{p},\{\op{x},\op{\rho}(t)\}]\;.\nonumber
\end{align}
Away from $\theta=\pi/4$, the CL($\theta$) master equation generally does not take Lindblad form, but remains a controlled time-local approximation within its regime of validity.
The $p$-$p$ coupling variant of the Caldeira-Leggett equations is obtained at $\theta=\pi/2$~\cite{Gelin2021, Venkatraman2024}.

The mixed coupling has a direct realization in superconducting circuit architectures:
in circuit quantum electrodynamics, the canonical variables are node flux $\Phi$ and charge $Q$, satisfying $[\Phi,Q]=i\hbar$, which play the roles of position and momentum~\cite{Krantz2019, Blais2021}.
Inductive coupling generates interactions of the form $\Phi\Phi_\nu$, corresponding to position-position coupling, while capacitive coupling yields $QQ_\nu$, corresponding to momentum-momentum coupling.
We summarize the resulting CL($\theta$) master equations in Table~\ref{tab:summary_MEs}. 

Eqs.~\eqref{eq:gCL_d} and~\eqref{eq:gCL_e} are not present in the CL($\theta$) model.
These terms originate exclusively from the dynamical dressing of the system-bath interaction by the nonlinear potential, Eq.~\eqref{eq:gCL_d}, and external drive, Eq.~\eqref{eq:gCL_e}, and have no counterpart in standard Caldeira-Leggett or Lindblad formulations. Specifically, Eq.~\eqref{eq:gCL_d}, describes how intrinsic nonlinearities generate amplitude-dependent dissipative processes, leading to nonlinear damping and modified fluctuations. 
While Eq.~\eqref{eq:gCL_e} describes how external driving mixes with the dissipative channel, producing dissipation-induced corrections to the effective drive amplitude and phase through the interplay of coherent driving and environmental fluctuations.

Importantly, these contributions render the dissipative dynamics amplitude dependent, leading to nonlinear damping and modified effective driving across different dynamical regimes of the system.
Moreover, these terms originate exclusively from the $p$-$p$ coupling channel.
In the absence of $p$-$p$ coupling, the dissipator reduces to the conventional Caldeira-Leggett form, and nonlinearities and driving remain confined to the unitary dynamics.
Momentum coupling therefore provides the microscopic mechanism through which system dynamics enter the dissipative sector.
At the semiclassical level, many of the resulting observable effects can be interpreted as amplitude-dependent damping and dissipation-induced renormalization of the effective drive.

The basis of our formalism for studying quantum master equations was previously applied only to static nonlinearities and established analogous terms~\cite{Duffus2016, Duffus2017, Duffus2018}. Here, we extend this generalization to incorporate arbitrary polynomial potentials and multi-tone driving.
The coupling between driving and dissipation is also consistent with previous studies of driven open quantum systems~\cite{Gulacsi2025, Bernazzani2025, Kolodyifmmodenelsenfiski2018} and with analogous results obtained in the Langevin framework~\cite{Ford1988, Cuccoli2001}.
More generally, effective Liouvillian approaches, such as higher-order van Vleck expansions, can generate additional mixing between nonlinearities, drivings and dissipation.
In circuit-QED implementations, all these effects correspond to capacitive coupling and provide a direct route to access and control these mechanisms experimentally.

Finally, we discuss the conditions under which the resulting dynamics remains physically well behaved.
The master equation derived here is trace preserving and Hermiticity preserving, but does not, in general, guarantee complete positivity away from specific parameter regimes.
At zero temperature, complete positivity is ensured only at the Lindblad point $\theta=\pi/4$, where the coefficient matrix of the master equation is positive semidefinite~\cite{BreuerPetruccione2002, Gaspard1999}.
At finite temperatures, the range of parameters yielding positive dynamics is extended (see Supplemental Material~\cite{Supplemental}).
Away from these regimes, the CL($\theta$) and gCL($\theta$) equations should be interpreted as effective Redfield-type descriptions.
They remain quantitatively reliable under physically relevant conditions such as weak system-bath coupling and sufficiently high temperatures, or for initial states that suppress short-time non-Markovian transients~\cite{Gaspard1999}.

\paragraph*{Summary: Generalized Caldeira-Leggett.} The generalized Caldeira-Leggett framework extends standard descriptions in two essential ways:
\begin{itemize}
    \item momentum-mediated coupling provides a continuous interpolation between position and momentum dissipation within a unified microscopic model, and
    \item intrinsic nonlinearities and external driving modify the dissipative dynamics through dynamical dressing of the system-bath interaction.
\end{itemize}
These mechanisms generate nonlinear damping together with dissipation-induced corrections to the effective drive, leading to observable signatures in both classical and quantum regimes.

From here on, we omit the explicit $\theta$ dependence and simply write CL and gCL.
\begin{table}[t]
\centering
\begin{tikzpicture}
\node[draw, rounded corners, inner sep=2pt] {
\begin{tabular}{ccc}
Interpolation angle $\theta$ & Coupling channel & Dissipative structure \\
\hline
$0$ & $x$-$x$ & \makecell{Caldeira-Leggett\\~\cite{BreuerPetruccione2002, Schlosshauer2008, Caldeira1983, Caldeira1983a}} \\
\rowcolor{rowgray}
$\pi/4$ & $x$-$x$ and $p$-$p$ & \makecell{Lindblad form\\~\cite{Duffus2016,Duffus2017,Duffus2018,Bai2005,Huang2022,Walls1970, bhattacharjee2024}
} \\
$\pi/2$ & $p$-$p$ & \makecell{Caldeira-Leggett\\~\cite{Gelin2021, Venkatraman2024, Cuccoli2001}} \\
\end{tabular}
};
\end{tikzpicture}
\caption{\textit{Interpolation structure of the CL($\theta$) master equation.}
The dissipative part of Eqs.~\eqref{eq:gCL_a} through~\eqref{eq:gCL_c} interpolates continuously between pure position coupling, pure momentum coupling, and a Lindblad structure at $\theta=\pi/4$.
}
\label{tab:summary_MEs}
\end{table}


\section{Application: Linearly driven dissipative Kerr oscillator}
\label{sec: application}

We illustrate the physical consequences of the gCL master equation~\eqref{eq: improved_Liouvillian} using the paradigmatic driven Kerr oscillator~\cite{eichler2018parametric}. 
This model allows us to isolate and experimentally interpret the distinct effects of nonlinear damping and dissipation-induced corrections to the effective drive predicted by Eq.~\eqref{eq: improved_Liouvillian}.
We set
\begin{equation}
    V_1(\op{x})=\frac{m^2\omega_0^2U}{3\hbar^2}\op{x}^4\;,
    \label{eq:def_Kerr_nonlin}
\end{equation}
and 
\begin{equation}
    V_2(\op{x},t)=F\cos(\omega t)\op{x}\;.
    \label{eq:def_lin_drive}
\end{equation}
where $U$ denotes the Kerr nonlinearity, and $F$ and $\omega$ are the drive amplitude and frequency.
As in the general case above, the damping rate $\gamma$ and the momentum coupling parameter $\theta$ characterize the system-bath interaction.
From here on, we set $m=\hbar=1$.

\subsection{Semiclassical nonlinear dissipation corrections}
\label{subsec: nonlin_damping_cl}
\begin{figure*}[t]
    \centering
    \includegraphics[width=1\linewidth]{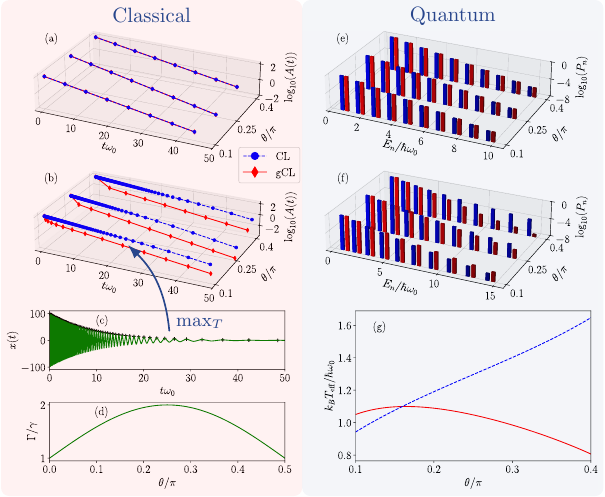}
    \caption{
\textit{Classical relaxation and quantum steady-state distributions illustrating the effect of nonlinear damping in the gCL model.}
(a) and (b) The envelope $A(t)$ of the semiclassical trajectories $x(t)$ [see (c)] for Kerr nonlinearity $U/\omega_0=0$ and $U/\omega_0=0.2$, respectively. There, the intrinsic nonlinearity induces a strongly amplitude-dependent decay from the nonlinear damping term. The CL model is plotted in blue colors and the gCL in red colors for different values of $\theta$. (d) The corresponding effective linear damping rate $\Gamma$ relative to $\gamma$ as a function of $\theta$. (e) and (f) The occupation probability $P_n = \langle \psi_n|\op{\rho}|\psi_n\rangle$ versus eigenenergy $E_n$ of the Kerr nonlinear oscillator for $U/\omega_0=0$ and $0.2$. 
These quantum distributions are inherently non-thermal, but can be fitted to a Bose-Einstein distribution to extract an effective temperature $T_{\rm eff}$.
(g) The extracted $T_{\rm eff}$ over the coupling angle $\theta$ for $U/\omega_0=0.2$. Throughout all panels, we apply the following global parameters: the bare frequency is $\omega_0=1$, the base damping rate is $\gamma/\omega_0=0.2$, and the momentum coupling angles are $\theta/\pi\in\{0.1,0.25,0.4\}$. For the quantum evaluations, we additionally set the thermal occupation to $n_{\rm th}=0.3$.
}
    \label{fig:dist_p_n}
\end{figure*}

We first isolate the effect of intrinsic nonlinearities on dissipation in Eq.~\eqref{eq:gCL_d} in the absence of external driving ($F=0$).
The semiclassical equation of motion for the oscillator position $x(t)=\langle \op{x}\rangle(t)$ reads~\cite{Supplemental}
\begin{align}
    \ddot{x}(t) = &- \omega_0^2 x(t)
    - \frac{4\omega_0^2 U}{3} x^3(t)
    - (1+\sin(2\theta))\gamma \,\dot{x}(t)
    \nonumber \\
    &- (1-\cos(2\theta))\,2\gamma U x^2(t)\dot{x}(t) \nonumber \\
    &-(\cos(\theta)+\sin(\theta))^2\sin(2\theta)\frac{\gamma^2}{2}x(t) \nonumber\\
    &-\sin(\theta)^2(1+\cos(2\theta)+\sin(2\theta))\frac{2\gamma^2U}{3}x^3(t)\;.
    \label{eq:sc_x}
\end{align}

The first line describes the standard dynamics of a damped nonlinear oscillator containing a harmonic restoring force, Kerr nonlinearity, and linear damping. The linear damping force is rescaled in dependence of $\theta$.
The second line is a nonlinear damping term proportional to $x^2(t)\dot{x}(t)$ arising from the nonlinear dissipative contribution in Eq.~\eqref{eq:gCL_d}.
The two final terms $\propto \gamma^2$ rescale both system frequency (in the CL and gCL case) and nonlinearity (only in the gCL case). 
All these additional contributions originate exclusively from momentum-momentum coupling and in the limit $\theta=0$ Eq.~\eqref{eq:sc_x} reduces to the standard classical equation of motion for a damped nonlinear oscillator, containing only linear damping, derived from a standard Caldeira-Leggett approach.
However, for simplicity, we focus on the nonlinear damping, which has no analogue in the standard Caldeira-Leggett treatment.

Standard Lindblad descriptions of weakly nonlinear systems do not capture such amplitude-dependent (nonlinear) damping. 
In experimental settings, nonlinear dissipation is often introduced phenomenologically via explicitly nonlinear system-bath couplings~\cite{Massignan2015, zaitsev2009nonlineardampingmicromechanicaloscillator,zaitsev2012nonlinear, Supplemental}.
In contrast, the present framework shows that nonlinear damping emerges naturally from a bilinear system-bath interaction once momentum coupling and intrinsic nonlinearities are retained.

To quantify the effect of the nonlinear damping, we analyze the ringdown dynamics~\cite{buks2006mass,Eichler2011,zaitsev2012nonlinear,polunin2016characterization,catalini2021modeling,mitman2023nonlinearities}, which provides a direct experimental probe of amplitude-dependent dissipation.
In such an experiment, the system is prepared in a high-amplitude state and evolved freely under the impact of damping. 
We simulate this process to extract the slowly varying envelope $A(t)$ of the trajectory $x(t)= A(t) \cos(\omega_0 t + \phi(t))$, where $\phi(t)$ is a slowly accumulating phase (see Fig.~\ref{fig:dist_p_n}(c)).
We then compare the nonlinear ringdown predicted by the full gCL dynamics and by the CL model for different coupling angles $\theta$.
Figures~\ref{fig:dist_p_n}(a) and (b) compare the cases $U/\omega_0=0$ and $U/\omega_0=0.2$ across different values of $\theta/\pi\in\{0.1,0.25,0.4\}$. 
In panel (a), we establish the linear baseline, where the dynamics follows a standard exponential envelope.
The relaxation occurs with an effective decay rate $\Gamma=(1+\sin(2\theta))\gamma$, reflecting the renormalization of linear damping by the momentum coupling channel, as shown in Fig.~\ref{fig:dist_p_n}(d).

Conversely, in Fig.~\ref{fig:dist_p_n}(b), we illustrate the impact of the intrinsic nonlinearity and resulting nonlinear damping in the gCL model. Since we prepare the system at high initial amplitudes, the nonlinear damping term dominates the ringdown at short times. 
Crucially, the early-time decay deviates from a simple exponential due to nonlinear damping.
As the amplitude decreases, the nonlinear contribution $x^2(t)$ becomes negligible and the dynamics crosses over to conventional linear damping with rate $\Gamma$.
Importantly, the standard CL framework does not capture this amplitude-dependent behavior.
At the semiclassical level, this effect can be interpreted as an amplitude-dependent renormalization of the effective damping.

\paragraph*{Summary: Classical nonlinear damping.}
The intrinsic nonlinearity of the system generates amplitude-dependent dissipation at the semiclassical level through momentum-mediated system-bath coupling.
This produces a two-stage relaxation: an initial fast decay governed by nonlinear damping, followed by a crossover to linear exponential relaxation.
Such behavior is absent in standard Caldeira-Leggett and Lindblad descriptions and can be directly probed via ringdown measurements.

\subsection{Quantum nonlinear dissipation corrections}
\label{subsec: nonlin_damping_qm}

In order to parametrize the effects of the nonlinear dissipation on the quantum system, we examine the steady-state. Even in the absence of external driving, intrinsic quantum fluctuations continuously excite the system, leading to a stationary distribution over the energy eigenstates.

We compute the steady-state occupation probabilities 
$P_n=\langle\psi_n|\op{\rho}|\psi_n\rangle$ for the Kerr eigenstates $|\psi_n\rangle$ with energies $E_n$, see Figs.~\ref{fig:dist_p_n}(e)-(g).
Figure~\ref{fig:dist_p_n}(e) corresponds to the linear case in Fig.~\ref{fig:dist_p_n}(a).
In the absence of nonlinearities, both the gCL and CL master equations yield identical results.
The population distribution exhibits exponential decay across equally spaced energy levels, allowing extraction of an effective temperature $T_{\rm eff}$ via a Bose-Einstein fit.
Here, $T_{\rm eff}$ serves as a phenomenological measure of the distribution width rather than a thermodynamic temperature.
Still, we choose the environment to exhibit a finite temperature, such that for both trace-preserving CL and gCL master equations, which however are not generally completely positive for all parameter regimes, the resulting steady-states remain well behaved, and the density matrix is completely positive.

Turning on the intrinsic nonlinearity in Fig.~\ref{fig:dist_p_n}(f) leads to three distinct effects:
(i) unequal energy level spacing,
(ii) strong dependence of the steady-state on the coupling angle $\theta$, and
(iii) a clear deviation between CL and gCL predictions.
While the CL model captures the modified spectrum, it lacks the corresponding nonlinear dissipative processes.
In contrast, the gCL framework includes both nonlinear damping and its associated fluctuations, redistributing the population across energy levels.

Because quantum noise rather than a purely thermal bath governs the stationary distribution, the populations are non-thermal, as seen in Fig.~\ref{fig:dist_p_n}(f).
Nevertheless, an effective temperature $T_{\rm eff}$ can be extracted as a phenomenological measure of the underlying fluctuations and width of the distributions, see Fig.~\ref{fig:dist_p_n}(g).

Increasing the $p$-$p$ coupling leads to a monotonic broadening of the steady-state distribution in the standard CL model.
In contrast, the gCL steady-state exhibits a nonmonotonic dependence on $\theta$.
The extracted $T_{\rm eff}$ initially increases with $\theta$ up to $\theta/\pi\approx0.16$ and decreases for larger angles.
Around this value, the gCL distribution becomes narrower than its CL counterpart, cf.~Fig.~\ref{fig:dist_p_n}(f).

This nonmonotonic behavior reflects a competition between nonlinear diffusion and nonlinear dissipation.
For small $\theta$, the diffusion term $[\op{x},[\op{x}^3,\op{\rho}]]$ dominates and broadens the distribution.
For larger $\theta$, the dissipative term $[\op{p},\{\op{x}^3,\op{\rho}\}]$ suppresses fluctuations and drives the system towards lower energies.
This competition leads to a qualitative reshaping of the steady-state distribution that is absent in standard descriptions.
This shows that nonlinear dissipation modifies both relaxation rates and the steady-state fluctuation distribution.

\paragraph*{Summary: Quantum steady-state.}
Nonlinear dissipation modifies both relaxation dynamics and steady-state fluctuations.
Within the gCL framework, nonlinear damping and its associated noise reshape the stationary distribution, producing non-thermal populations and a nonmonotonic dependence on $\theta$.
These features provide experimentally accessible signatures that distinguish gCL from standard Caldeira-Leggett models.


\subsection{Dissipation-induced drive and resonance asymmetry}
\label{subsec: driven_system}

In this section we consider how the dynamics of the system are modified by the dissipation-induced drive correction generated by Eq.~\eqref{eq:gCL_e}.
To this end, we consider a linearly driven harmonic oscillator ($U=0$, $F\neq0$), where nonlinear effects are absent and the dissipation-induced corrections to the effective drive can be analyzed separately.

For this system, in the quantum gCL master equation Eq.~\eqref{eq:gCL_e} is seen to generate an additional linear time-periodic momentum drive. Consequently, the semiclassical equations of motion read
\begin{align}
    \ddot{x}(t) \approx &- \omega_0^2 x(t)
    - (1+\sin(2\theta))\gamma \dot{x}(t) -F\cos(\omega t)
    \nonumber \\
    &
    +(1-\cos(2\theta))\frac{\gamma F\omega\sin(\omega t)}{2\omega_0^2}\nonumber\\
    &-(\cos(\theta)+\sin(\theta))^2\sin(2\theta)\frac{\gamma^2}{2}x(t) \nonumber\\
    &-\sin(\theta)^2(1+\cos(2\theta)+\sin(2\theta))\frac{\gamma^2F\cos(\omega t)}{2\omega_0^2}\;.
    \label{eq:sc_x_2}
\end{align}
Since the equation of motion is linear, it holds both in the quantum and classical regime. We again separate the standard dynamics of a driven damped harmonic oscillator in the first line of Eq.~\eqref{eq:sc_x_2} from the remaining terms. The second line represents a dissipation-induced correction to the effective drive, appearing as an additional phase-shifted driving force.
At the semiclassical level, this contribution can be interpreted as a renormalization of the drive amplitude and phase through the dissipative channel.
The two final lines $\propto \gamma^2$ in Eq.~\eqref{eq:sc_x_2} rescale both system frequency (in the CL and gCL case) and drive strength (only in the gCL case).
All these additional contributions are absent for pure $x$-$x$ coupling and therefore provide a direct dynamical signature of dissipation-induced drive mediated by momentum coupling.
We focus in particular on the phase-shifted driving force induced by dissipation.

To quantify the observable consequences of this additional driving channel, we analyze the steady-state response of the system.
\begin{figure}[t]
\centering
\includegraphics[width=1.\linewidth]{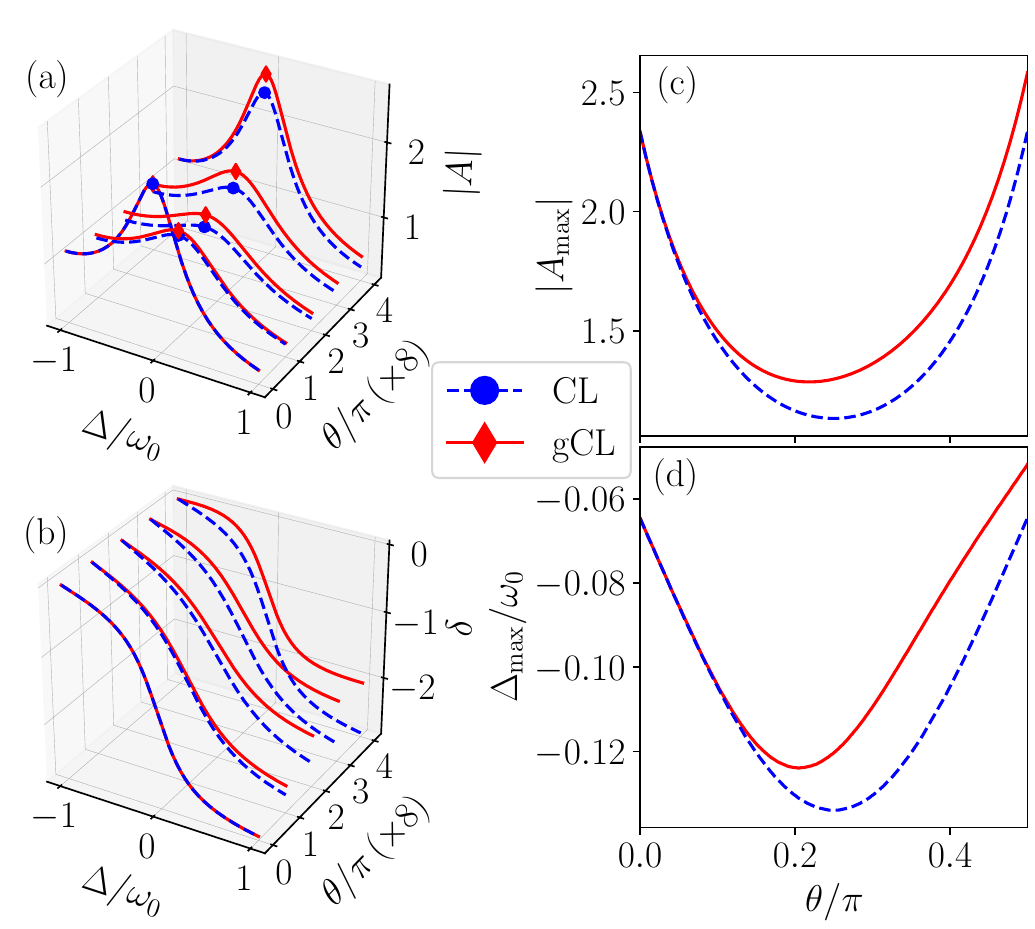}
\caption{\textit{Response fingerprints of dissipative drive corrections.}
(a) The response amplitude $|A|$ as a function of detuning $\Delta$ for the linear driving regime. The maximum points are explicitly marked. There, the CL response (blue lines) maintains strict symmetry around $\theta=\pi/4$, whereas the gCL response (red lines) exhibits a pronounced asymmetry. (b) The phase shift $\delta$ as a function of detuning. 
We again observe that the CL result is symmetric around $\theta=\pi/4$, whereas this symmetry is broken in the gCL case.
Both results exhibit a shift in the system's natural frequency, the phase shifted drive in the gCL result however leads to a smaller total absolute phase shift with increasing detuning. (c) and (d) The amplitude $A_{\rm max}$ and detuning $\Delta_{\rm max}$ of the resonance maximum as a function of $\theta$, respectively. Both panels illustrate explicit symmetry breaking in the gCL model. Throughout all panels, we apply the following global boundary conditions: $F_q/\omega_0=0.4$, $\gamma/\omega_0=0.5$, and $\theta/\pi\in\{0,1/8,1/4,3/8,1/2\}$ (increasing with line intensity).}
\label{fig:drive_response}
\end{figure}
We consider the steady-state solution of the equation of motion Eq.~\eqref{eq:sc_x_2}.
We use the ansatz $x(t)=A\cos(\omega t+\delta)$, where $A$ is the response amplitude and $\delta$ the phase shift. 
For convenience, we define the rescaled drive strength $F_q=\hbar F/2\sqrt{2\hbar\omega_0 m}$. 
We then characterize the system's response by mapping the amplitude $|A|$ against the detuning $\Delta=\omega-\omega_0$, see Fig.~\ref{fig:drive_response}(a).  

Both the CL and gCL descriptions yield Lorentzian-like resonance line shapes, reflecting the linearity of the underlying dynamics.
However, their dependence on the coupling angle $\theta$ differs qualitatively.
The response amplitude in both models exhibits a minimum near the Lindblad point $\theta=\pi/4$. 
Within the standard CL description, this dependence is strictly symmetric around $\theta=\pi/4$. 
This symmetry follows as the CL dynamics only renormalizes the damping coefficient and frequency in Eq.~\eqref{eq:sc_x_2}, both of which depend symmetrically on $\theta$.
In contrast, the gCL response displays a pronounced asymmetry in Fig.~\ref{fig:drive_response}(a). 
This asymmetry originates directly from the dissipation-induced, phase-shifted driving term in Eq.~\eqref{eq:gCL_e}.
Since this term introduces a contribution in the orthogonal quadrature $\sin(\omega t)$, it modifies the effective forcing rather than simply renormalizing system parameters.
As a result, the response amplitude is no longer constrained to be symmetric around $\theta=\pi/4$, but is systematically larger than the CL result for larger $\theta$.

We also consider the phase shift $\delta$ as a function of detuning $\Delta$ in Fig.~\ref{fig:drive_response}(b) and observe the same symmetry behavior.
Furthermore, dissipation-induced frequency renormalization is visible in both the CL and gCL results through a shift of the $\delta=-\pi/2$ point away from $\Delta=0$. However, the additional dissipation-induced drive in the gCL master equation leads to systematically smaller phase shifts due to its earlier onset relative to the original drive.
This demonstrates that the dissipation-induced drive correction cannot be reduced to a symmetric renormalization of the effective damping and resonance frequency, but instead acts as an additional phase-shifted driving contribution.
The resulting resonance asymmetry in detuning is reminiscent of Fano-type interference phenomena, where the observable response arises from interference between direct and indirect excitation pathways~\cite{Fano_1961}. Here, the momentum-dependent dissipation-induced correction generates an additional phase-shifted bath-mediated driving channel that interferes with the coherent drive.

Both Figs.~\ref{fig:drive_response}(a) and (b) show corresponding signatures of the dissipation-induced correction to the effective drive. To analyze the role of the momentum damping in greater detail, we examine the resonance maxima. 
We extract the peak amplitude $A_{\rm max}$ and corresponding detuning $\Delta_{\rm max}=\omega_{\rm max}-\omega_0$ as a function of the mixing angle $\theta$, see Figs.~\ref{fig:drive_response}(c) and (d).


In the CL model, both quantities remain symmetric around $\theta=\pi/4$, with maximal amplitude and minimal detuning at this point.
In contrast, the gCL response displays a pronounced asymmetry.
For small $\theta$, the response approaches the CL limit, as expected when momentum coupling is weak.
For larger values of $\theta$, the resonance amplitudes and frequency shifts increase in both models, but with clearly distinct functional dependencies.
As a result, the resonance minimum shifts to $\theta\approx0.2\pi$, and the corresponding detuning exceeds the CL prediction.

These asymmetric features provide a direct experimental signature of the additional time-dependent drive arising from the interplay between coherent driving and dissipation.
Measuring the $\theta$ dependence of resonance amplitudes and frequency shifts provides a direct macroscopic probe to distinguish between standard Caldeira-Leggett and generalized dissipative dynamics~\cite{Ochs_2022, EichlerZilberberg2023}.

\paragraph*{Summary: Dissipation-induced drive.}
Momentum-mediated system-bath coupling generates an additional phase-shifted driving force through the interplay between coherent linear driving and environmental fluctuations.
This produces a characteristic symmetry breaking in the resonance response as a function of $\theta$ and a Fano-type asymmetry in detuning, which is absent in standard Caldeira-Leggett models.
These features provide a clear and experimentally accessible signature of dissipative drive renormalization induced by momentum-mediated coupling.


\begin{figure}[t]
\centering
\includegraphics[width=1.0\linewidth]{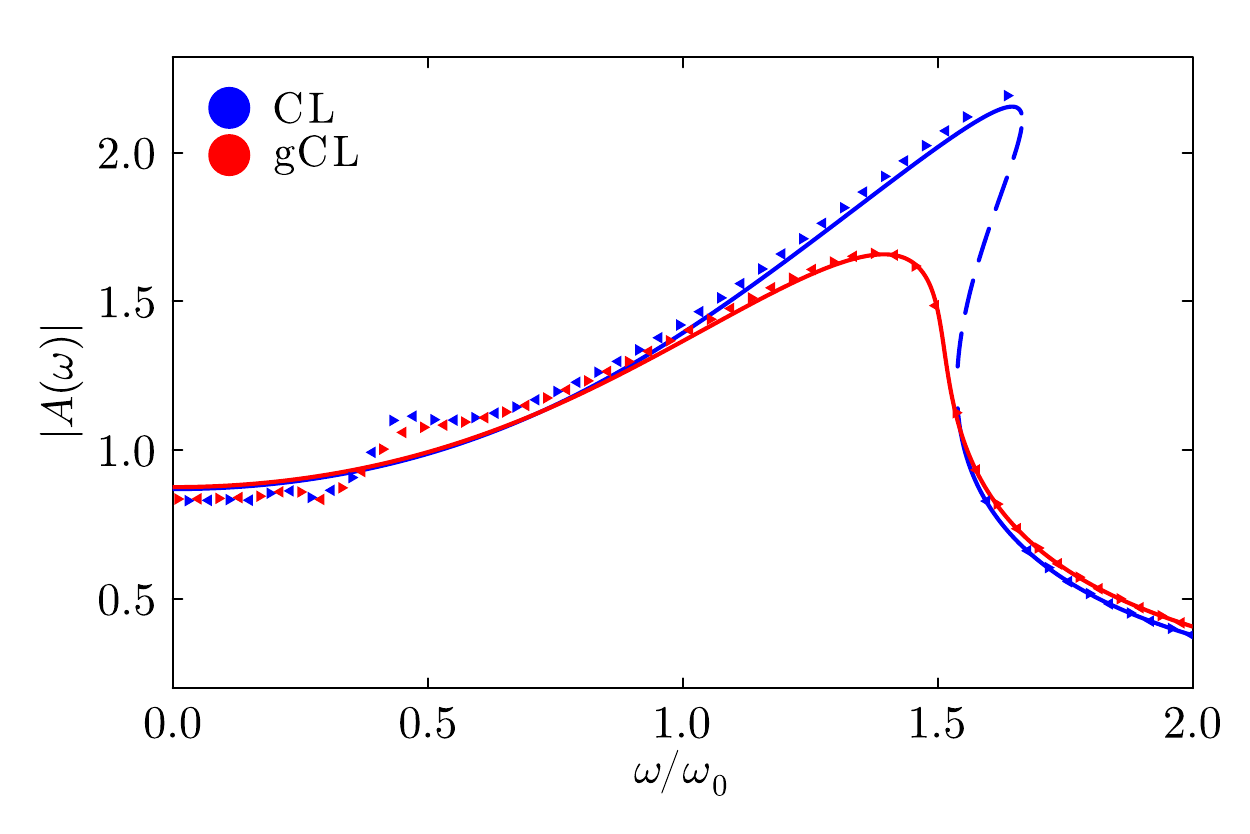}
\caption{\textit{Bistability suppression by nonlinear dissipation.}
Response amplitude $|A(\omega)|$ of a linearly driven Kerr oscillator as a function of driving frequency $\omega$, shown for the CL and gCL master equations.
Forward and backward frequency sweeps, indicated by the direction of the markers, reveal bistability in the CL case, which is eliminated in the gCL model.
In the generalized framework, amplitude-dependent damping together with the dissipation-induced drive correction suppress the bistable regime, replacing the discontinuous switching between metastable branches by a smooth crossover.
The dissipation-induced drive correction additionally shifts the resonance response.
Solid and dashed lines show response functions obtained within a rotating wave approximation, corresponding to stable and unstable solutions, respectively.
Parameters are $\omega_0=1$, $\gamma/\omega_0=0.2$, $U/\omega_0=3/8$, $F_q/\omega_0=0.4$, and $\theta=0.4\pi$\;.
}
\label{fig:response_stability}
\end{figure}

\subsection{Interplay of nonlinear and dissipation-induced drive}
\label{subsec: MPR}

We now consider regimes in which nonlinear damping and dissipation-induced drive corrections act simultaneously.
In this regime, Eqs.~\eqref{eq:gCL_d} and~\eqref{eq:gCL_e} jointly modify both the steady-state response and quantum fluctuations.
Their combined effect modifies the steady-state response and reduces the overall phase-space fluctuation scale while largely preserving the resonance structure.
At the semiclassical level, the suppression of large-amplitude excitations is dominated by the amplitude-dependent nonlinear damping, whereas the dissipation-induced drive correction primarily modifies the effective forcing through an additional phase-shifted quadrature component.
Physically, this reflects that the dissipative dynamics inherits the operator structure of the drive, which can modify the distribution of fluctuations across phase space.
The dominant effect is a reduction of the total fluctuation scale $\nu_{\mathrm{geo}}$.
For linear driving, modifications of the anisotropy $R$ remain subleading, whereas for higher-order driving they become significant and reflect a redistribution across quadratures.

We illustrate this behavior for two representative cases:
(i) a Kerr oscillator with linear driving, and
(ii) a Kerr oscillator with two-photon driving.

\subsubsection{Kerr oscillator with linear driving: stabilization and fluctuation suppression}

Consider a quantum oscillator subject to both Kerr nonlinearity, Eq.~\eqref{eq:def_Kerr_nonlin}, and linear driving, Eq.~\eqref{eq:def_lin_drive}.
In this regime, amplitude-dependent nonlinear damping and dissipation-induced drive corrections act simultaneously, and the dynamics is governed by the full gCL master equation in Eq.~\eqref{eq: improved_Liouvillian}.

At the classical level, we analyze the frequency-dependent response amplitude $|A(\omega)|$.
Within the standard Caldeira-Leggett description, the system exhibits the well-known bistability of a driven Kerr resonator~\cite{Drummond1980, Casteels2017}.
In contrast, the gCL framework strongly suppresses this bistability through enhanced amplitude-dependent damping and modified effective driving, while largely preserving the resonance structure, as shown in Fig.~\ref{fig:response_stability}.
 The bistability region is shifted in parameter-space.
This stabilization originates from the combined action of the phase-shifted dissipation-induced driving term and amplitude-dependent damping.
The former modifies the effective forcing, while the latter enhances damping at large amplitudes.
Together, they reshape the response and suppress the resonant amplification.
The resulting modification is already captured at the level of a rotating-wave approximation, corresponding to the leading order of a Floquet or Krylov-Bogolyubov expansion~\cite{HB_jl, EichlerZilberberg2023}, and is fully consistent with the numerical solutions in Fig.~\ref{fig:response_stability}.

\begin{figure}[t]
\centering
\includegraphics[width=\linewidth]{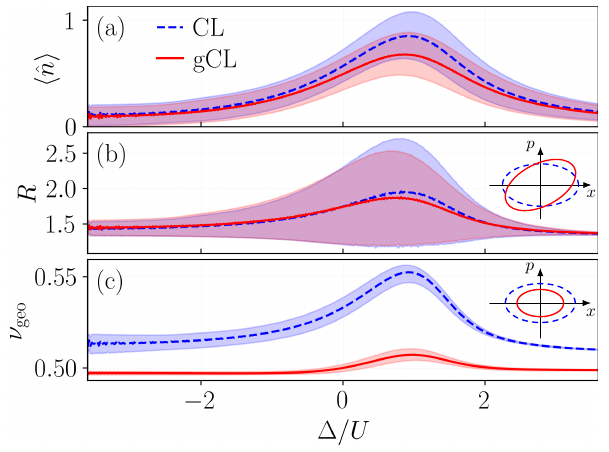}
\caption{
\textit{Suppression of fluctuations in the linearly driven Kerr oscillator.}
(a) Mean occupation $\langle \hat n\rangle$ as a function of normalized detuning $\Delta/U$, comparing the CL (dashed blue) and gCL (solid red) master equations.
(b) Anisotropy of quadrature fluctuations, quantified by $R$.
(c) Total phase-space fluctuation level, measured by $\nu_{\mathrm{geo}}$.
The gCL dynamics suppresses the resonant response while largely preserving the overall resonance structure.
The dominant effect is a strong reduction of $\nu_{\mathrm{geo}}$ across the resonant region, while the anisotropy $R$ is only weakly modified.
Shaded regions indicate the $10$-$90\%$ range of residual long-time oscillations, while solid and dashed lines denote time-averaged values.
Insets illustrate the phase-space covariance ellipse: panel (b) emphasizes its aspect ratio, quantified by $R$, while panel (c) emphasizes its overall size, quantified by $\nu_{\mathrm{geo}}$.
Parameters are $\omega_0=1$, $\gamma/\omega_0=0.3$, $U/\omega_0=0.2$, $F_q/\omega_0=0.3$ and $\theta=\pi/4$.
}
\label{fig:FplusU}
\end{figure}

We now turn to the quantum steady-state and characterize both the excitation level and fluctuations.
The mean occupation is defined as
\begin{equation}
\op{n}=\frac{\omega_0}{2}\op{x}^2+\frac{1}{2\omega_0}\op{p}^2-\frac{1}{2}\op{\mathbb{1}},
\label{eq:occ_operator}
\end{equation}
while fluctuations are quantified via the symmetrized covariance matrix
\begin{equation}
\Sigma_{ij} = \langle \op{r}_i \op{r}_j + \op{r}_j \op{r}_i \rangle/2 - \langle \op{r}_i \rangle \langle \op{r}_j \rangle,
\qquad
\op{\mathbf r} = (\op{x}, \op{p}) .
\end{equation}
Its eigenvalues, denoted by $\nu_{\min}$ and $\nu_{\max}$, define the principal quadrature fluctuations. 
From these, we construct the total fluctuation scale $\nu_{\mathrm{geo}} = \sqrt{\nu_{\min}\nu_{\max}}$ and the anisotropy ratio $R = \nu_{\max}/\nu_{\min}$.
All observables are evaluated in the long-time regime.
Residual micromotion due to the periodic drive is accounted for by reporting time-averaged values together with their fluctuation range.

The resulting detuning dependence is shown in Fig.~\ref{fig:FplusU}.
Both the mean occupation and the fluctuation scale $\nu_{\mathrm{geo}}$ exhibit a broad resonance around $\Delta/U\sim 1$.
Within the CL description, this resonance is associated with enhanced occupation and fluctuations.
In contrast, the gCL dynamics suppresses both quantities while preserving the overall resonance structure.

This effect is most clearly visible in panel (c), where $\nu_{\mathrm{geo}}$ is strongly reduced across the resonant region, demonstrating a direct suppression of the fluctuation scale.
At the same time, panel (b) shows only weak modifications of $R$, indicating that the suppression is approximately isotropic, with only subleading changes to the quadrature structure.

Taken together, these results show that amplitude-dependent nonlinear damping together with dissipation-induced drive corrections reduce the fluctuation scale $\nu_{\mathrm{geo}}$, while leaving the anisotropy $R$ largely unchanged and preserving the overall resonance profile.

\begin{figure}[t]
\centering
\includegraphics[width=\linewidth]{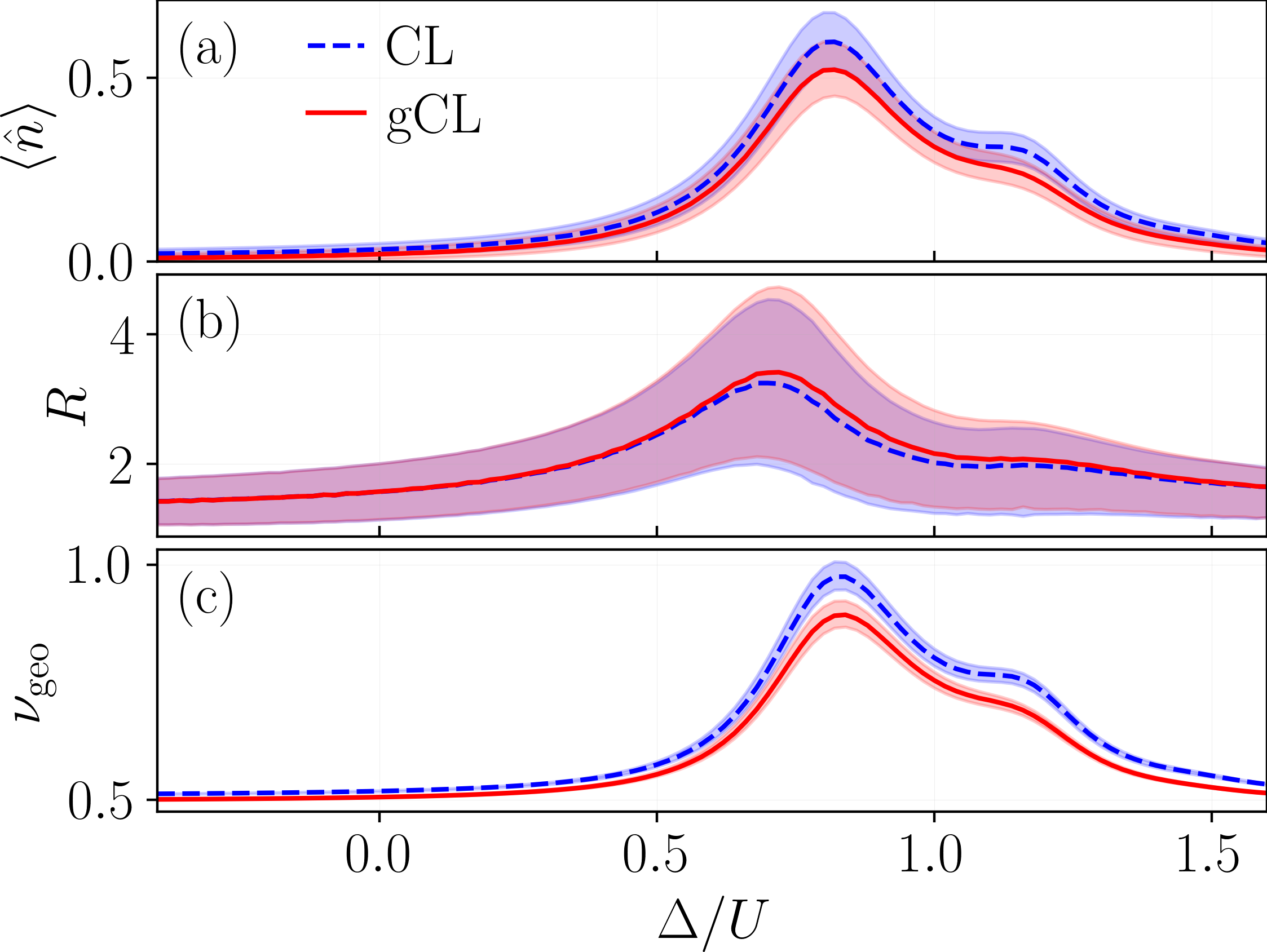}
\caption{
\textit{Reshaping of parametrically driven resonances by nonlinear dissipation.}
(a) Mean occupation $\langle \hat n\rangle$ as a function of normalized detuning $\Delta/U$ for the two-photon driven Kerr oscillator, comparing CL (dashed blue) and gCL (solid red) dynamics.
(b) Anisotropy of quadrature fluctuations, quantified by $R$.
(c) Total phase-space fluctuation level, measured by $\nu_{\mathrm{geo}}$.
The parametrically driven system exhibits a structured resonance profile with a primary peak and a secondary shoulder.
The gCL dynamics preserves this overall structure while suppressing the resonant response.
The total fluctuation scale $\nu_{\mathrm{geo}}$ is reduced across the resonant region, while the anisotropy $R$ is modified nontrivially, indicating a redistribution of fluctuations across quadratures.
Shaded regions indicate the $10$-$90\%$ range of residual long-time oscillations, while solid and dashed lines denote time-averaged values.
Parameters are $\omega_0=1$, $\gamma/\omega_0=0.03$, $U/\omega_0=0.2$, $G/\omega_0=0.1$ and $\theta=\pi/4$ .
}
\label{fig:GplusU}
\end{figure}

\subsubsection{Kerr oscillator with two-photon driving: reshaping of parametrically driven resonances}

We next consider a Kerr oscillator subject to two-photon driving,
\begin{align}
    V_2(\op x, t)=F_2\cos(2\omega t)\op x^2.
\end{align}
This system provides a complementary probe of the generalized dissipative dynamics.
Unlike linear driving, the two-photon drive generates nontrivial operator structures within the dissipative dynamics, which directly affect the distribution of fluctuations in phase space.
With this flavor of driving, the time-dependent dissipative contributions in Eq.~\eqref{eq:gCL_e} renormalize both position drift $\propto [\op p,\{\op x,\op\rho\}]$ and momentum diffusion $\propto [\op x,[\op x,\op\rho]]$ channels, leading to a qualitatively modified fluctuation profile.

Figure~\ref{fig:GplusU} shows the resulting behavior in terms of the mean occupation and covariance-based measures of fluctuations.
The response exhibits a multi-feature structure with a pronounced primary peak and a secondary shoulder on the high-detuning side, reflecting the multi-resonant nature of the parametrically driven nonlinear oscillator.
Within the CL description, both features are associated with enhanced occupation and increased phase-space fluctuations.
In contrast, the gCL dynamics preserves the overall structure while suppressing both quantities across the resonant region.

This suppression is non-uniform.
The secondary feature at higher detuning is more strongly attenuated, indicating that excitations associated with this part of the resonance are more efficiently damped.
At the same time, the total fluctuation scale $\nu_{\mathrm{geo}}$ is reduced across both features, corresponding to a reduction of $\nu_{\mathrm{geo}}$.
In contrast to the linearly driven case, this reduction is accompanied by pronounced modifications of $R$, where now the squeezing given by the gCL model is larger than for the CL one, indicating that the suppression is not isotropic but involves a redistribution of fluctuations across quadratures.

The anisotropy of fluctuations, quantified by $R$, exhibits a clear reshaping compared to the CL dynamics.
This behavior reflects the operator structure of the dissipator in Eq.~\eqref{eq:gCL_e}, where the two-photon drive introduces contributions that couple quadratically to the system quadratures.
As a result, the generalized dissipative dynamics not only reduces the fluctuation scale $\nu_{\mathrm{geo}}$ but also redistributes fluctuations across quadratures.

Overall, the parametrically driven case is characterized by:
(i) a global suppression of fluctuations, and
(ii) a drive-dependent redistribution in phase space, leading to a selective attenuation of resonant features.


\paragraph*{Summary: driven Kerr resonators.}

Across both driving protocols, amplitude-dependent nonlinear damping and dissipation-induced drive corrections suppress excitations and phase-space fluctuations while largely preserving the overall resonance structure.
For linear driving, the dominant effect is an approximately isotropic suppression of fluctuations with bistability removed by nonlinear dissipation.
For two-photon driving, the nontrivial operator structure of the dissipator reshapes the fluctuation anisotropy and produces a non-uniform suppression of resonant features.
These effects originate from the same momentum-mediated dissipative mechanism encoded in Eqs.~\eqref{eq:gCL_d} and~\eqref{eq:gCL_e}, which generate nonlinear damping and dissipation-induced corrections to the effective drive beyond standard Caldeira-Leggett dynamics.


\FloatBarrier
\section{Conclusions and outlook}
\label{sec: conclusion}

We have developed a generalized Caldeira-Leggett framework for nonlinear, driven quantum oscillators with mixed position- and momentum-dependent system-bath coupling.
We retain the full nonlinear and time-dependent system dynamics in the system-bath evolution, resulting in a master equation that describes dynamically dressed dissipation beyond standard Lindblad and purely $x$-$x$ Caldeira-Leggett approaches.
This generates nonlinear and drive-dependent dissipative channels, leading to amplitude-dependent damping, dissipation-induced corrections to the effective drive, and strongly modified fluctuation distributions.
As a consequence, nonlinearities and external driving modify not only the coherent evolution, but also the dissipative structure of the open-system dynamics itself.
Taken together, our results identify momentum-mediated system-bath coupling as the microscopic channel through which nonlinear dynamics and external driving reshape the dissipative sector.
For driven Kerr oscillators, this leads to profound modification of regimes where classical bistability manifests, Fano-type asymmetric resonance responses, and reduced phase-space fluctuations while largely preserving the underlying resonance structure.

Looking ahead, our framework can be extended to non-Markovian environments~\cite{Gulacsi2025, Gaspard1999, Hu1992}, multimode systems, and interacting oscillator arrays.
Its predictions can be tested experimentally through measurements of switching dynamics, resonance asymmetry, fluctuation statistics, and quadrature-resolved observables in superconducting circuits, nanomechanical resonators, and cavity-QED architectures.
More broadly, our results show that dissipation in driven nonlinear quantum systems cannot generally be regarded as a passive linear background, but instead becomes dynamically structured by nonlinear motion and coherent driving.
Accounting for this dynamically dressed dissipation is important for describing and engineering quantum devices operating beyond linear-response regimes.

\section{Acknowledgments}
The authors acknowledge stimulating discussions with Michel H. Devoret and Mark I. Dykman during the Workshop on Parametric Phenomena 2025 in Konstanz. 
We thank D. K. J. Bone{\ss} and B. Gulácsi for valuable comments on the manuscript. We acknowledge funding from the Deutsche Forschungsgemeinschaft (DFG) through project numbers 449653034 and 545605411, as well as through the research unit FOR5688 (project number 521530974) and via SFB 1432 (project number 425217212). We also acknowledge support from the Swiss National Science Foundation (SNSF) through NCCR SPIN and through Sinergia Grant No. CRSII5\_206008/1.
\bibliographystyle{KilianStyle}
\bibliography{Effective_liouv.bib}

\end{document}